\documentstyle[12pt,epsf]{article}
\rmfamily
\def\Z0{${\em Z^0\/}$}

\def\into{\rightarrow}
\def\ovl{\overline}
\def\r#1 {$^{#1}$}

\newcommand{\et}{{\rm E}_{\scriptscriptstyle\rm T}}

\newcommand{\met}{\mbox{$\protect \raisebox{.3ex}{$\not$}\et$}}

\newcommand{\ppbar}{p\overline{p}}
\newcommand{\qqbar}{q\overline{q}}
\newcommand{\ttbar}{t\overline{t}}
\newcommand{\bbbar}{b\overline{b}}
\newcommand{\ccbar}{c\overline{c}}

\def \mc {\multicolumn}

\begin{document}
\begin{flushright}
{\large \bf FERMILAB-CONF-99/246-E \\ 
September 1, 1999 }
\vspace{0.5in}
\end{flushright}
\begin{center}
\begin{large}
{\bf SINGLE TOP PRODUCTION AND TOP PROPERTIES AT THE TEVATRON }\\ 
\end{large}
\vspace{0.5in}
\begin{large}
{Angela Barbaro-Galtieri} \\
 {\it Lawrence Berkeley National Laboratory} \\  ~ \\
 { For the CDF and  ${\rm D\!\not\!\!O}$ ~Collaborations }\\ 
\end{large}
\end{center}
\vspace{1.0in}

\begin{abstract}
 We summarize recent measurements made at the Tevatron Collider using top 
 event candidates. Cross section and mass measurements are discussed
 in a separate contribution to these Proceedings. Here we report on studies 
 of the top P$_T$ distribution in $\ttbar$ production and studies of single 
 top production. Properties of top decays examined are: 
 BF($t \into Wb$)/BF($t \into Wq$), helicity amplitudes of W's from 
 top decays and correlations of $\ttbar$ decay products. Searches for new 
 physics in rare top decays and a search for a state X $\into \ttbar$ are 
 also  reported.
\end{abstract}

\newpage
\section{Introduction}
\label{intro}
%
The top quark is produced in $\ppbar$ collision mostly via $\ttbar$ pair
production. At $\sqrt{s}$=1.8 TeV top pairs are produced 90\% of the time 
via  $\qqbar$ diagrams, the remainder via  {\it{gg}} diagrams. 
Within the Standard Model (SM) top decays into W+b $\sim $ 100\% of the time.
The W decays are as follows:
\begin{center}
\begin{tabular}{|c|c|c|c|c|c|} \hline
$W^+\rightarrow$ & $e^+\nu$ & $\mu^+\nu$ & $\tau^+ \nu$ & $u\ovl d$ &
 $c\ovl s$ \\ \hline
Br & 1/9 & 1/9 & 1/9 & 3/9 & 3/9 \\ \hline
\end{tabular}
\end{center}
%
For $\ttbar$ production the event topologies are:
\begin{itemize}
 \item Dilepton channel ($\ell \nu \ell \nu \bbbar$): events for which 
     both W's decay into $e ~or~ \mu$. This is expected to occur in
     a fraction of 4/81 events, i.e., $\sim 5$\% of the final
     states. Also  $e ~or~ \mu$ plus $\tau$ is expected to occur in $\sim 5$\%
     of the events.
 \item Lepton+jets channel ($\ell \nu \qqbar \bbbar$): events in which
    one W decays into $e ~or~ \mu$, the other into a quark pair. This occurs
   with a fraction 24/81, i.e. in $\sim 30$\% of the events.
 \item All-hadronic channel ($\qqbar \qqbar \bbbar$), events in which  
    both W's decay into quark pairs. This occurs with a fraction 36/81,
    i.e. in $\sim 44$\% of the events.
 \end{itemize}
 About one hundred top events have been detected so far at the Fermilab Tevatron
 ($\sqrt s$ = 1.8 TeV) in the ${\rm D\!\not\!\!O}$
 and CDF experiments in Run I. 
\begin{center}
\begin{tabular} {|l|l||c|c|c||c|c|} \hline
  & Sample& \mc{3}{c||} {CDF}  & \mc{2}{c|} {${\rm D\!\not\!\!O}$} \\ \hline       
  &       & $a \times \epsilon$ & Events& Background &Events& Background \\
 \hline
 1 & Dileptons(ee,e$\mu$,$\mu \mu$) & 0.0074 &  9  & 2.4 $\pm$ 0.5 
                                             &  5  & 1.4 $\pm$ 0.4 \\
 2 & Dileptons($e(\mu) + \tau$)     & 0.0013 &  4  & 2.5 $\pm$ 0.4 
                                             & n.a.& n.a.          \\
 3 & $e \nu$                        & n.a. & n.a.& n.a. & 4  & 1.2 $\pm$ 0.4 \\ 
 4 & $\ell$+$\ge$ 3jets (SVX tag)   & 0.037  & 34  & 9.2 $\pm$ 1.5 
                                             & n.a.& n.a. \\
 5 & $\ell$+$\ge$ 3jets (SLT tag)   & 0.017  & 40  & 22.6 $\pm$ 2.8 
                                             & 11  & 2.4 $\pm$ 0.5 \\
 6 & $\ell$ + 4jets(no-tag)         &  n.a.  & 46 & 25.9 $\pm$ 6.5
                                             & n.a.& n.a        \\
 7 & $\ell$ + 4jets(loose jet4)     &  n.a.  & 163& 108 $\pm$ 13  
                                             & n.a.& n.a   \\
 8 & $\ell$ + jets (topological)    & n.a. & n.a.& n.a.& 19 & 8.7 $\pm$ 1.7 \\
 9 & All-hadr (1 SVX)               & 0.044  & 187 & 142 $\pm$ 12
                                             & n.a.& n.a  \\
10 & All jets                       & n.a. & n.a.& n.a.& 41 & 24.8 $\pm$ 2.4 \\
    \hline
\end{tabular}
\end{center}
%
%
Details of the CDF events are as follows.
$a\times \epsilon$ = acceptance $\times$ efficiency. 
E$_T$(lepton) is required to be $>$ 20~GeV and the missing transverse
energy is required to be $\met >$ 20~GeV. A jet is defined as
a calorimeter cluster with E$_T >$ 15 GeV in a cone of radius R=0.4. Events
in the lepton ($e, \mu$)+jets channel are divided into sub-samples 
depending upon whether they include or not a jet tagged 
 as $b$ jet. Tags are established by the presence of a displaced
vertex (SVX) or of a soft lepton from a $b$ semileptonic 
decay (SLT). The SLT sample includes many events with an SVX tag.
 The $\ell$ + 4jets (loose jet4) sample allows a fourth cluster 
 with E$_T >$ 8 GeV in the event. This sample includes samples 4-6.
For CDF the All-hadronic sample requires one of the $\ge$ 6 jets
to be tagged as a $b$ jet by the SVX algorithm. The ${\rm D\!\not\!\!O}$ 
sample selection is
somewhat different and can be found in the many references given below.

\begin{figure}[htbp]
\centerline{
\epsfysize 7.4cm
\epsffile[100 145 615 680] {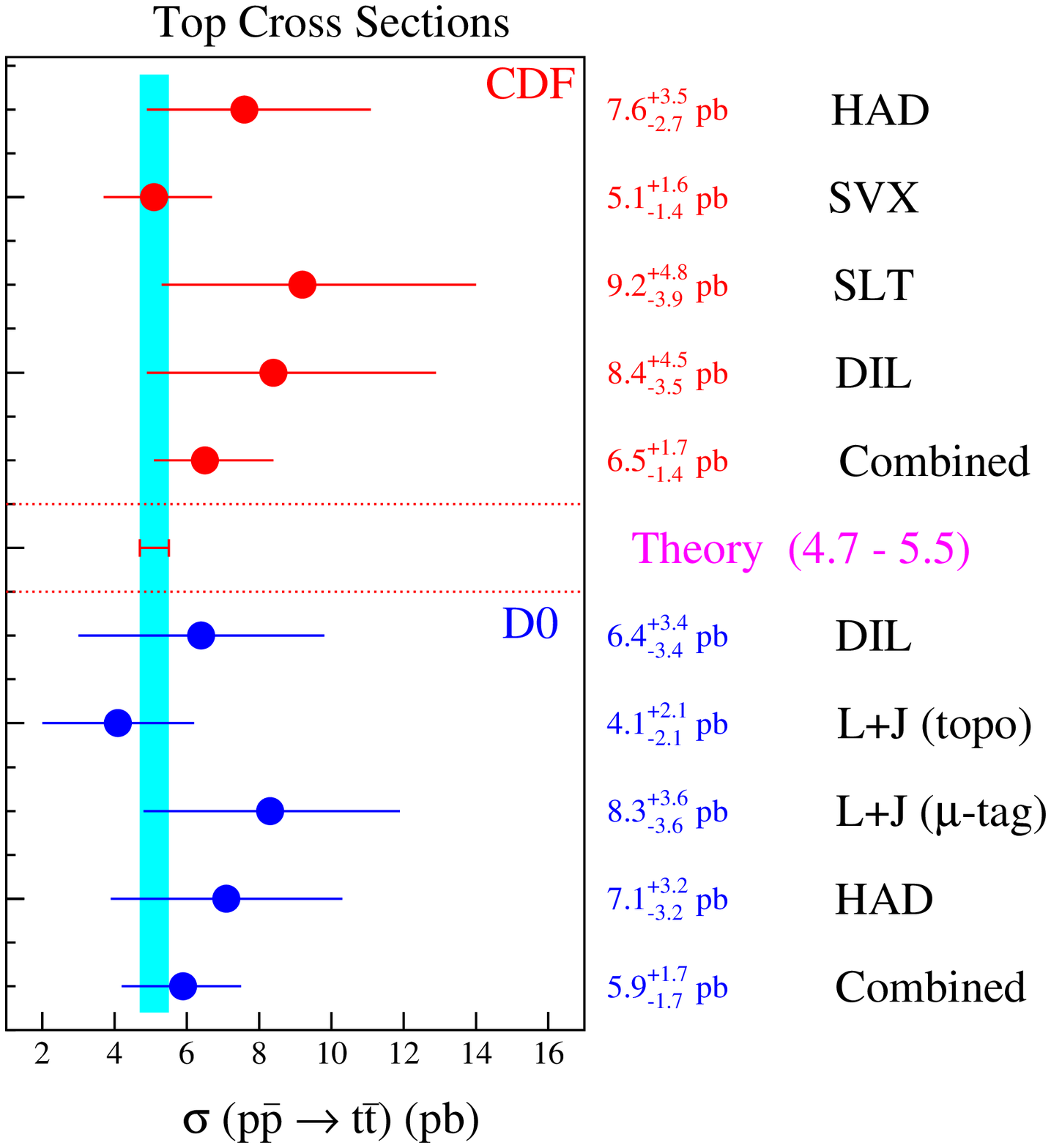}
 \epsfysize 7.8cm
 \epsffile[85 160 510 650] {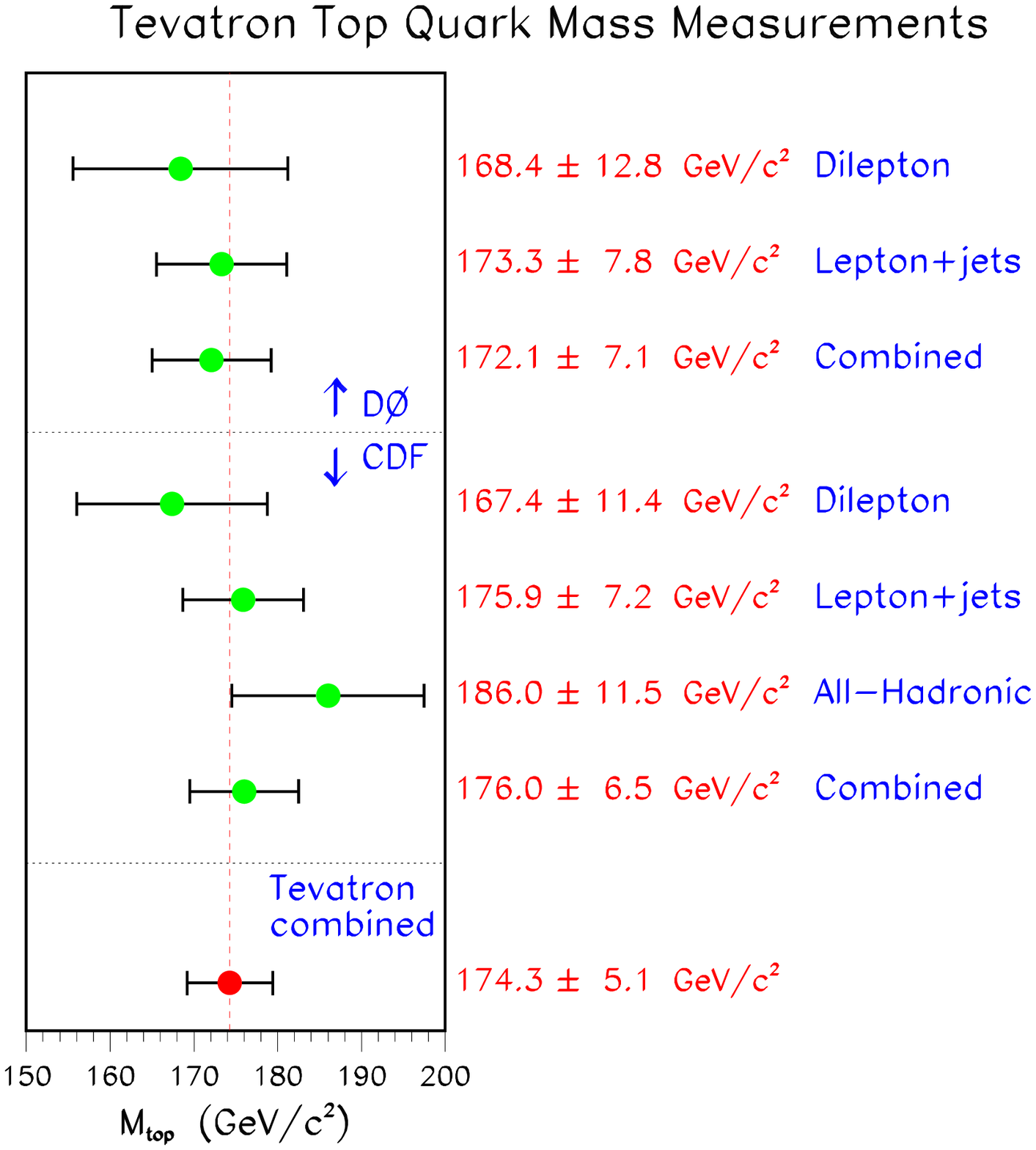 }}
\caption{\label{sig-mass} Cross section (left) and mass (right) 
 measurements by CDF and ${\rm D\!\not\!\!O}$.}
\end{figure}

\begin{figure}[htbp]
 \centerline{
 \epsfysize=3.0in
 \epsffile[80 220 525 618] {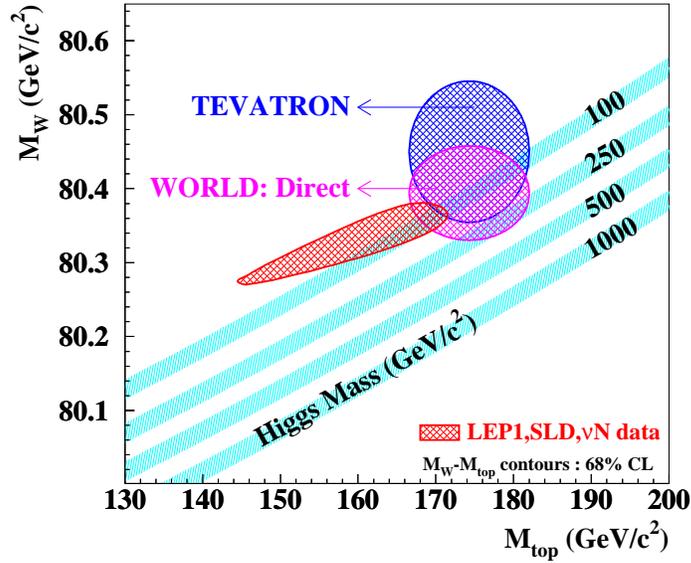 }}
 \caption{\label{mwmt_mh} Relation between the W and top masses for some values
    of the Higgs mass. The cross hatched area to the left represents the
    allowed region from an overall Standard Model fit. The other two regions
    are obtained from direct measurements of M$_W$ and M$_{top}$. } 
\end{figure}

Cross section measurements by both CDF~\cite{sigma_CDF} and 
${\rm D\!\not\!\!O}$\cite{sigma_D0} have been made and are 
shown in Figure~\ref{sig-mass}.
${\rm D\!\not\!\!O}$ has presented an updated cross section at this 
Conference~\cite{top_wimp,d0_sign}.
CDF has updated its cross section measurements since this Conference. At this
writing the combined cross sections for each experiment are:  
\begin{eqnarray}
 \sigma_{\ttbar} =  6.5^{+1.7}_{-1.4} ~pb~~~@ ~M_{top}=175~GeV/c^2 ~~~~ CDF \\
 \sigma_{\ttbar} =  5.9^{+1.8}_{-1.7} ~pb~~~@ ~M_{top}=172~GeV/c^2 ~~~~ 
             ~{\rm D\!\not\!\!O}~
\end{eqnarray}
The top mass value is specified, because the acceptance is a weakly dependent
function of mass. 
Direct top mass measurements were also made by CDF~\cite{cdf_mass} and 
${\rm D\!\not\!\!O}$~\cite{d0_mass} in
several channels and are shown in Figure~\ref{sig-mass}.  
The combined  mass from the CDF and ${\rm D\!\not\!\!O}$ experiments is~\cite{top_wimp}:
\begin{eqnarray}
 M_{top} =  174.3 \pm 5.1 ~GeV/c^2 
\end{eqnarray}
%
%
Measurements of the top mass and the W mass test the Standard Model 
beyond tree level.
The W mass values from the Tevatron~\cite{tev_wmass} and  
LEP II~\cite{lep_wmass}  are:
\begin{center}
 CDF + ${\rm D\!\not\!\!O}$ average M$_W$ = 80.450 $\pm$ 0.063 ~GeV/c$^2$~~~~  \\
 LEP II average~~~~M$_W$ = 80.375 $\pm$ 0.078 ~GeV/c$^2$  \\
 World average~~~~~M$_W$ = 80.420 $\pm$ 0.049 ~GeV/c$^2$
\end{center}
Figure~\ref{mwmt_mh} shows the relation between M$_W$ and M$_{top}$ 
for a constant value of M$_{Higgs}$ as expected by the Standard Model (SM).
Within the SM precise measurements of  M$_W$ and M$_{top}$ impose a 
constraint on the Higgs mass. The allowed region from present Tevatron 
measurements is shown as well as that obtained by using the world average
for M$_W$. The allowed region obtained from an overall fit of many
measurements by LEP experiments, the SLD detector at the SLC 
and by neutrino experiments, is shown as the narrow oval region on the left.
The contours are one standard deviation uncertainties and are in good 
agreement with each other.

%

In Run II (2 fb$^{-1}$ of integrated luminosity, 2 TeV $\ppbar$ collisions 
and upgraded detectors) the Tevatron experiments 
expect up to a factor 40 more data in most channels, hence better
determination of the top and W masses. These measurements
will be dominated by systematic uncertainties.

Here we report on other studies done at the Tevatron to investigate
agreement between top production measurements and SM expectation,
as well as preliminary searches for new physics. In Section 2 we describe 
a study of the P$_T$ distribution of top quarks in $\ttbar$ production. 
In Section 3
we investigate single top production. In Section 4 we report on decay
properties of the top quark. Finally, in Section 5 we investigate beyond the
SM phenomena, i.e., rare top decays and a search for a Topcolor
 state
X $\into \ttbar$. All of these studies can benefit from more statistics and are
really done in preparation for Run II data.

\section{ Top Pair Production   }

 At the Tevatron ($\sqrt s$ = 1.8 TeV) the $\ttbar$ production 
cross section from QCD calculations~\cite{top_sig} 
is expected to be in the range 4.75 $pb$ to 5.5 $pb$ 
for M$_{top}$ = 175 GeV/c$^2$.
As shown earlier, measurements done by CDF and ${\rm D\!\not\!\!O}$ are in reasonable 
agreement with this 
prediction within the measurement errors. Are the features of the 
events compatible with what is predicted
by QCD top pair production? Comparisons of the lepton+jets data samples
with Monte Carlo predictions for $\ttbar$ production have been made.
 CDF has used many variables in the comparison and very good 
agreement between data and expectations has been found~\cite{CDF_kin}.
Similar studies were done by ${\rm D\!\not\!\!O}$ 
with the same conclusions~\cite{D0_ljetm}.

\subsection{ Top P$_T$ in $\ttbar$ Production (CDF)} 
 \label{top_pt}
\begin{figure}[htbp]
 \centerline{
 \epsfysize=2.0in
 \epsffile[15 15 290 205] {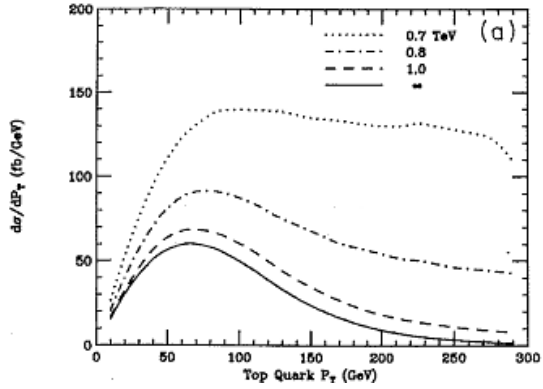 }}
 \caption{\label{zp_pt} Top P$_T$ distribution expected for a color-singlet
        state, Z', decaying into $\ttbar$ in the Topcolor model of Hill and 
         Park[12]. }
\end{figure}

A detailed study of the top P$_T$ distribution has been done by CDF.
This variable has been chosen because in some models it is more sensitive
to new physics than the total cross section and large deviation from QCD
are expected. 
For example, the Topcolor model of Hill and Park predicts color octet
(and/or singlet) vectors associated with top condensation that enhance or
depress the top production cross section~\cite{top_color}. A particular 
case within this model predicts a narrow Z' that 
decays into a $\ttbar$ pair. It predicts the top 
P$_T$ distributions shown in 
Figure~\ref{zp_pt} for different values of the new scale $\Lambda$.

\begin{figure}[htbp]
 \centerline{
 \epsfysize=5.5cm
 \epsffile[95 260 480 490] {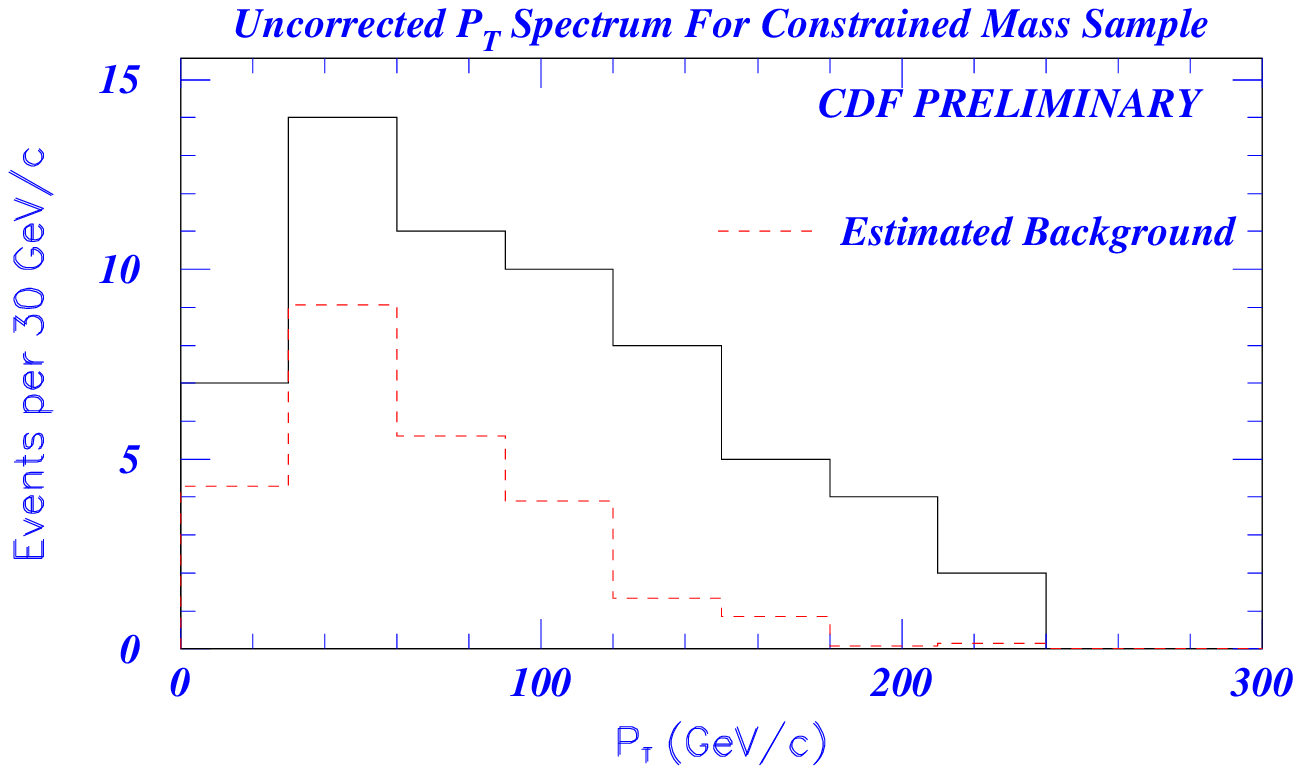 }
 \epsfysize=6.8cm
  \epsffile[-20 0 405 425] {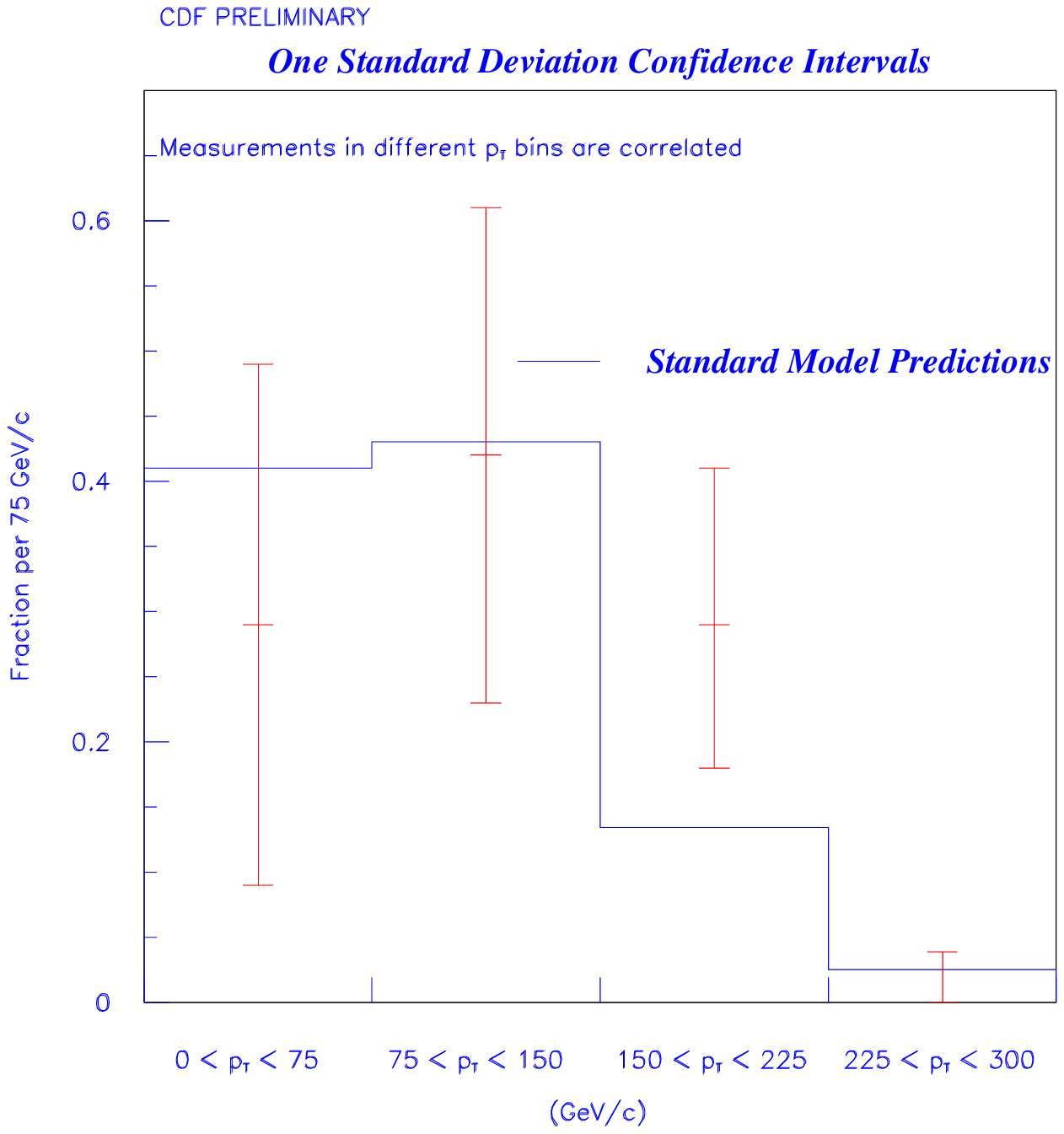 }}
 \caption{\label{tpt_cdf} Left: observed top P$_T$ distribution from 
   CDF reconstructed $\ttbar$ events in the W+jets sample. 
   The estimated background distribution is also shown.
   Right: the ``true'' top P$_T$ distribution
   for top events, obtained after unfolding the experimental resolution
   and acceptance correction.} 
\end{figure}

CDF uses the lepton + jet sample to which the top mass kinematic fitting 
for reconstructing the event is applied~\cite{ljet_mass}, 
with M$_{top}$ fixed at 175
GeV/$c^2$. After the requirement of a good fit, 61 events 
remain with 
a background of 24.6 $\pm$ 5.8 events. The distribution in P$_T$ for
these events is shown in Figure~\ref{tpt_cdf} (left plot), along
with the estimated background distribution. 
%
To extract the P$_T$ spectrum of produced top quarks, CDF unfolds for 
resolution
smearing and corrects for acceptance as a function of top P$_T$. 
This is done in four bins of ``true'' top P$_T$. 
An unbinned likelihood estimator, which uses the response functions 
for $\ttbar$ and background, is then employed to find the top contribution.
Response functions for the four bins are obtained from the
HERWIG top Monte Carlo plus detector simulation. The VECBOS Monte 
Carlo~\cite{vecbos}
is used for background. An 
iterative technique is used to minimize the assumptions made 
about the shape of 
the P$_T$ spectrum within each bin. The final ``true'' top 
P$_T$ distribution at production is shown in Figure~\ref{tpt_cdf} (right plot).
The error bars in each bin include statistical and
systematic uncertainties. The different 
sources of systematics are:
M$_{top}$ value, jet E$_T$ scale, gluon radiation, background shape,
acceptance. 
The results for each bin are shown in Table~\ref{tpt_4bins}.
\begin{table} [h]
\begin{center}
\caption{\label{tpt_4bins}''True'' top P$_T$ distribution measured by CDF }
\begin{tabular} {|c|l|} \hline
  P$_T$ bin           & Measured Fraction of Top quarks \\   \hline
 ~~0 $<P_T<$ ~75 GeV/c & ~~0.29~$^{+0.18}_{-0.18}$(stat)~~$^{+0.08}_{-0.08}$(syst)
   \\
 ~75 $<P_T<$ 150 GeV/c & ~~0.42~$^{+0.18}_{-0.18}$(stat)~~$^{+0.05}_{-0.07}$(syst)
    \\
 150 $<P_T<$ 225 GeV/c & ~~0.29~$^{+0.12}_{-0.10}$(stat)~~$^{+0.06}_{-0.05}$(syst)
     \\
 225 $<P_T<$ 300 GeV/c & 0.000~$^{+0.035}_{-0.000}$(stat)~$^{+0.019}_{-0.000}$(syst)\\ \hline
\end{tabular}
\end{center}
\end{table}

The unfolded 
distribution is in good agreement with QCD $\ttbar$ production expectation,
i.e., with the standard model prediction.
The upper  limit for R$_4$, i.e., fraction of events 
with top P$_T$ in the 225-300 GeV/c bin, is calculated to be 
R$_4 <$ 0.114 at 95\% CL.
It seems in disagreement with a flat top P$_T$ distribution, like the one
shown in Figure~\ref{zp_pt} with a scale $\Lambda$ = 0.7 TeV. 
Clearly more statistics is
needed to exclude or prove any new model. This is a study that 
can be carried out with  Run II data.

\section { Single top production (CDF) }

\begin{figure}[htbp]
 \centerline{
 \epsfysize=3.4in
   \epsffile[70 170 380 680]{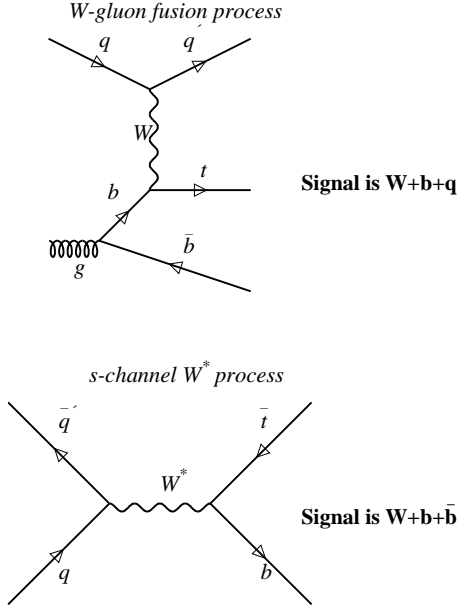 }}
\caption{\label{sigt_dia} Single top production diagrams.}
\end{figure}

Top quarks can be produced along with a $b$ quark through electro-weak 
interactions. Although the cross section is smaller than that for top pair
production, it is important to pursue this study because single top 
production 
can be used to measure the CKM matrix element V$_{tb}$. 
The process is 
 sensitive to new physics at the t-W-b vertex, like contributions from a
W' or flavor changing neutral currents (FCNC). At the Tevatron two
different diagrams   are expected to contribute (Figure~\ref{sigt_dia}).
They have different cross sections and 
require different methods of analysis.

We report here on analyses being done by the CDF Collaboration.

For W-g fusion $\sigma$ = 1.7 $\pm$ 0.3 $pb$ ~(See~\cite{W_g}).
The q' jet is expected to be energetic, whereas  
the $b$ jet at the lower vertex is soft. Since 
top pair production contributes appreciably  
in W+3jet events ($\sigma_{\ttbar}=5.1~pb$), only W+ 
2 jets events are used (E$_{T_3}<$15 GeV), and   
only one tagged $b$ jet is allowed. The  
sample used is then $\ell$ + 2jets (1 SVX tag). 
For the W$^*$ channel $\sigma$ = 0.73 $\pm$ 0.10 ~$pb$ 
(See~\cite{W_star}). Here the $b$ jet at the lower vertex 
is energetic and two $b$ jets are expected in 
the event. The sample used is then $\ell$+2jets  
with $\ge$ 1 SVX tag. Backgrounds come from   
QCD processes with tagged jets (W$\bbbar$, W$\ccbar$,
W$c$, W+2jets mistags, and $\ttbar$). 

\subsection{ W-gluon fusion process}
 
\begin{figure}[htbp]
 \centerline{
 \epsfysize=2.8in
 \epsffile[45 160 540 680]{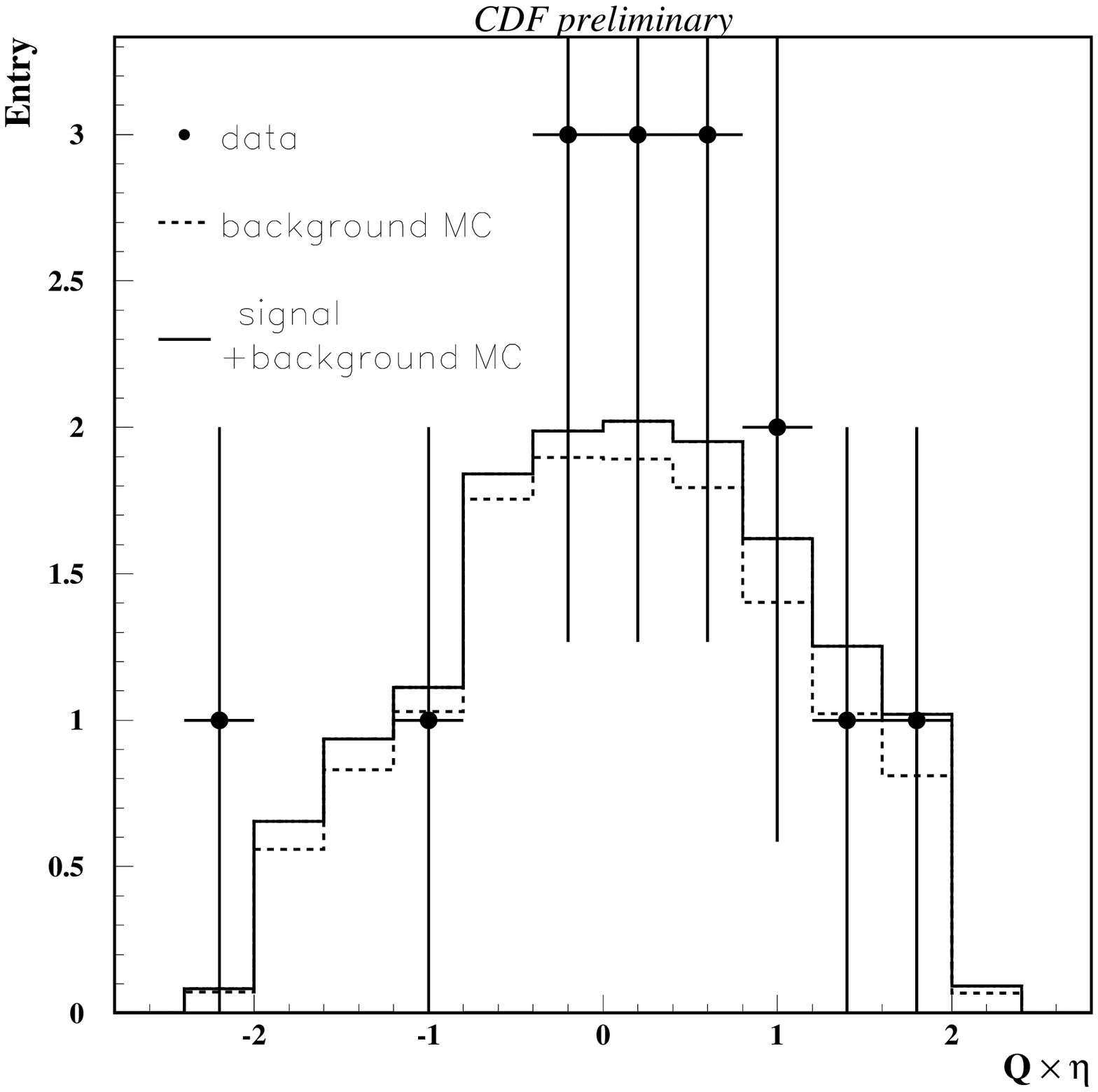 }
 \epsfysize=2.8in
 \epsffile[30 170 540 675]{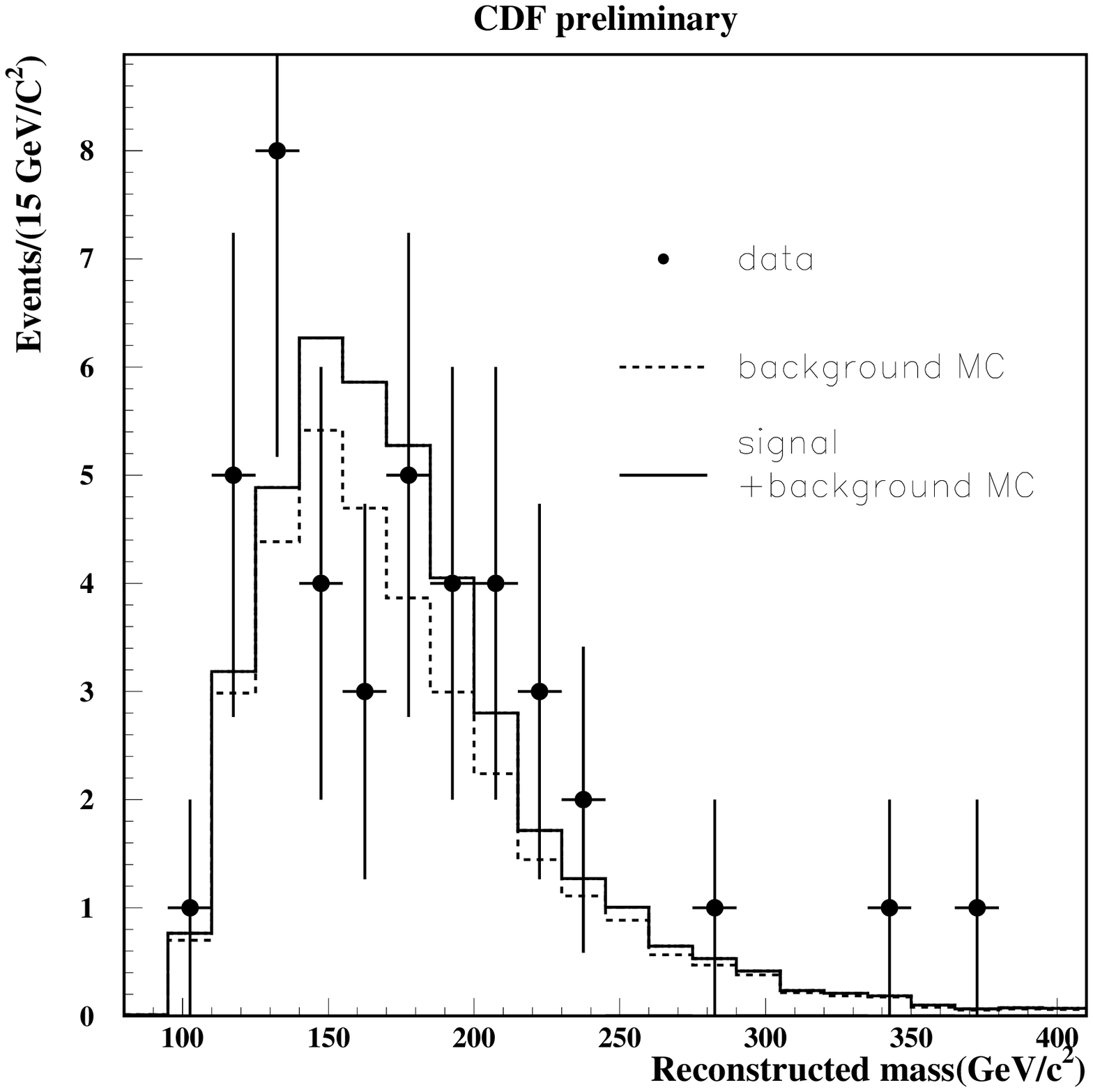 }} 
 \caption{\label{sing_top} Left: the Q$\times \eta$ distribution for the 
  W+2 jets sample used for W-gluon fusion study. Q is the charge of the lepton,
  $\eta$ is the pseudorapidity of the untagged jet. Right: the  M($\ell \nu b$)
  distribution for the W+2 jet sample used to study the W$^*$ process. }
\end{figure}

Backgrounds  for this process are the same as for the W* process.
The requirement that the $b$ jet at the lower vertex be soft,
reduces the acceptance by
a factor 2, while reducing the $\ttbar$ background by 85\%.
To reduce the QCD background from other sources, CDF requires 
145 $<$ M($\ell \nu b$) $<$ 205 GeV/c$^2$.
Expected and observed number of events are shown below.
\begin{center}
 \begin{tabular}{|l|c|c|c|c|c|}  \hline
 Source & QCD         & W*          & $\ttbar$    & W-g & Found \\ \hline
Events & 10.2$\pm$1.9 & 0.5$\pm$0.1 & 2.2$\pm$0.6 & 1.2$\pm$0.3 & 15 \\
\hline
\end{tabular}
\end{center}  
Monte Carlo studies find that the charge of the top quark, 
hence of the lepton (Q),
is correlated with the pseudorapidity, $\eta$, of the untagged jet 
($\ell^+$ has 
a jet with  positive $\eta$, $\ell^-$ has a jet with negative $\eta$). 
Therefore, the variable Q$\times \eta$ is used to extract an estimate of
the signal because signal events
tend to be asymmetric whereas the background is symmetric. The
distributions obtained from the data and those expected from background and 
background+signal are shown in Figure~\ref{sing_top}.

 A likelihood method is used to estimate the single top fraction. It uses
templates for W$\bbbar$, W + jets, $\ttbar$, and the other processes. The 
major systematic uncertainties included are from: Monte Carlo signal generator,
initial and final state radiation, jet energy scale, luminosity and 
$b$-tagging efficiency. 
The results of this procedure is  $\sigma_{W-g}$ = 
 1.9$^{+5.6}_{-4.5}$ $pb$, which results in the limit:
\begin{eqnarray} 
  \sigma_{W-g} < 15.4 ~pb ~~~~at~ 95\% ~CL 
\end{eqnarray}

\subsection{ W$^*$ process} 

Only events with $\ge$ 1 SVX  $b$-tagged jets are included in this study.
The expected and observed number of events are shown below. 
\begin{center}
 \begin{tabular}{|l|c|c|c|c|c|}  \hline
 Source &    QCD     &   $\ttbar$    & W-glu   &      W$^*$ & Found \\ \hline
Events & 24.0$\pm$4.5 & 5.7$\pm$1.3 & 1.6$\pm$0.3 & 1.0$\pm$0.3 & 42 \\
\hline
\end{tabular}
\end{center}  
To study the signal the variable of choice is M($\ell \nu b$), calculated with
the assumption that the $b$ with larger $\eta$ is
associated with $\ell^+$ in W$^+$ production (the opposite for W$^-$).
Monte Carlo studies show this choice to be correct 64\% of the time. 
Figure~\ref{sing_top} shows this distribution (right hand plot).

A maximum likelihood method is used to estimate individual contribution 
to the sample. Both statistical and systematic uncertainties are taken 
into account.
With fixed $\ttbar$ and W-g fusion cross sections, the analysis gives
$\sigma_{W^*}$ = 4.9$^{+5.5}_{-4.9} pb$. 
This gives an upper limit:
\begin{eqnarray} 
  \sigma_{W^*} < 15.8 ~pb ~~~~~at~95\%~~CL  
\end{eqnarray}

 \subsection{ Run II expectation} 
Limits obtained in Run I are an order of magnitude larger than the expected
cross sections. Is there a chance to obtain a measurement in Run II
( 2 fb$^{-1}$ of data)? Improvements expected are as follows:
\begin{itemize}
 \item Factor 20 in luminosity,
   Tevatron energy up to 2.0 TeV. At this energy the  top cross 
      sections are larger by 40\%, 
       whereas the rest of the QCD background increases only by 12\%.
 \item Detector improvements expected are: 1.)
  lepton acceptance increases, $e$ by 33\%, $\mu$ by 12\% and 2.)
     SVX $b$ tagging efficiency increases by about 50\%.
\end{itemize}
The expected number of events are:
\begin{center}
 \begin{tabular}{|l|c|c|c|c|}  \hline
 Source & ~~QCD~~ & ~~ $\ttbar$~~ & ~W-glu~   & ~~ W$^*$~~  \\ \hline
Events  & 170     &    80         &   50      & 15  \\
\hline
\end{tabular}
\end{center}
Additional handles (total event transverse energy and different E$_T$
threshold for jet 1) can be used to reduce backgrounds. It is likely that
more than 2 fb$^{-1}$ of data will be needed.

\section{ Top Decay Studies}
 
 Assuming that the top quark has isospin 1/2 and that it forms an SU(2)
 doublet with the bottom quark, the Standard Model has a definite prediction
on the allowed decay modes: BF($t \into Wb$) $\sim$100\%. Also this decay 
should exhibit the behavior expected by the V-A weak interaction, and 
precise predictions about the subsequent decays of the W 
can be made.
In this section we present three studies on details of the top decay
products.

\subsection{ Branching Fraction for $t\rightarrow Wb$ and $|V_{tb}|$ (CDF) }

The branching fraction for $t\rightarrow Wb$ is $\sim 100\%$ in the
        Standard Model.  A value significantly different from this could
        signal new physics. 
The expectation for decays into quarks other then $b$ 
is very low.  The BF($t\rightarrow Wb$) depends upon the value of
      $|V_{tb}|$, which  is constrained by the other elements of 
the CKM matrix  to be  0.9991 $< |V_{tb}| <$ 0.9994 when the assumptions of 
three generations and unitarity are made~\cite{pdg}.

CDF, using good $b$ tagging capabilities due to the presence of a silicon
detector close to the beam pipe,  measures the ratio
\begin{eqnarray}
 R_b = {BF(t\rightarrow Wb) \over BF(t\rightarrow Wq)}  
\end{eqnarray}
In the Standard Model the CKM matrix (3 generations and unitarity) makes a 
prediction of the value of  $R_b$ via the following relation:
\begin{eqnarray}
 R_b = {V_{tb}^2
               \over V_{td}^2 + V_{ts}^2 + V_{tb}^2 }
\end{eqnarray}
The above value of V$_{tb}$  suggests that R$_b$ is very close to one.

CDF uses a few independent data samples to measure  R$_b$. Events with:
  \begin{itemize}
     \item 0, 1 or 2 $b$-tagged jets in the $\ell$+4jets(loose jet4) sample.
     \item 0, 1 or 2 $b$-tagged jets in the Dilepton sample.
  \end{itemize}
The number of events in each category depends upon the
fraction of $b$'s present in top decays. The observed events are:
\begin{center}
\begin{tabular}{|l|c|c|c|c|c|}
\hline
Sample, ~~~~~~~~~~Tags   & &  0         &   1-SLT  & 1-SVX & 2-SVX \\ \hline
$\ell$+4jets(loose jet4)&observed & 126        &  14      & 18    &  5    \\
$\ell$+4jets(loose jet4)&backgr. & 107$\pm$10 & 7.2$\pm$1.6 &2.9$^{+2.3}_{-1.6}$
                         &  0.2$\pm$0.1    \\  \hline
Dileptons         &observed &  6   &  n.a.    &  3    &  0    \\
Dileptons         & backgr. & 2.3$\pm$0.5 & n.a. & 0.10 $\pm$0.04 & n.a. \\
 \hline
\end{tabular}
\end{center}
Backgrounds for each of these categories are estimated and included 
in a maximum likelihood estimator used to  measure
R$_b$. The result is:  
\begin{eqnarray}
  R_b = 0.93^{+0.31}_{-0.23} ~~~CDF ~~ 
\end{eqnarray}
Assuming unitarity CDF calculates V$_{tb}$:
\begin{eqnarray}
  |V_{tb}| = 0.96^{+0.16}_{-0.12}~~ or~~ |V_{tb}| ~> ~0.74 ~(95\% \rm{~CL}) 
\end{eqnarray}
Note that if the SM with three generations were not correct, the measurement 
of  $R_b$ could not be used to calculate  $|V_{tb}|$. The calculation of
$|V_{tb}|$ done here provides a check of the SM. The
single top production discussed in Section 3 would be a model independent
way to measure $V_{tb}$.

\subsection{ W Boson Helicity in Top Decays (CDF) }

The Standard Model has an exact prediction (in leading order) of 
      the W helicity
      states in the top rest frame. Since the top quark decays
      before it hadronizes the spin information at production is
      preserved. Because of the large top mass we expect:
\begin{eqnarray}
   F_0 = {{{{1} \over {2}} m_t^2/M_W^2} \over {1+{{1} \over {2}}
          m_t^2/M_W^2}}  
\end{eqnarray}
     For M$_{top}$ = 175 GeV/c$^2$ ~F$_0$ = 0.70. Also 
   F$_+$ = 0.0. 
Here F$_0$ is the fraction of longitudinal W bosons (H$_W$ =0)
 and F$_+$ is the fraction of right handed W bosons (H$_W$ = +1). 
Large deviations from these predictions would indicate that
top has non-SM couplings.

To test this prediction the relevant information can be obtained from decay 
angular distributions
of leptons in the top rest frame. This requires full reconstruction
of the event, with its uncertainties due to combinatorics. 
CDF uses the observed lepton P$_T$ distribution, which also carries
this information and does not require event reconstruction. 
This allows the use of all leptons from top decay candidates even if the 
kinematic fit did not succeed, thus providing a larger statistical
sample. This method, however, is more sensitive to the Monte Carlo model 
used, especially the background model.  

\begin{figure}[phtb]
 \centerline{
 \epsfysize=3.4in
 \epsffile[33 150 520 696] {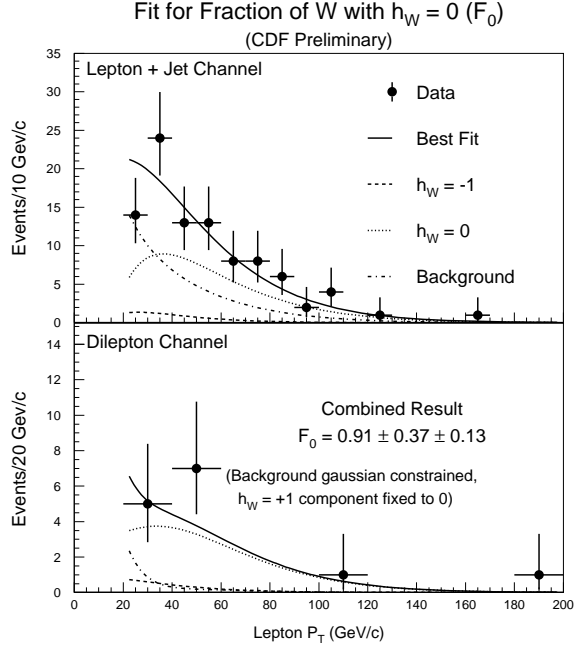 }}
 \caption{\label{whel_fit} Observed P$_T$ distributions of the
    lepton in $\ell$ + jets and dilepton top candidate samples. 
    The response functions for the background and top are shown. The solid
    curves show the results of a likelihood estimator of F$_0$. }
\end{figure}

CDF uses events from the Dilepton (only $e\mu$) and the lepton+jets (SVX 
and SLT tags and no-tag) samples to perform this 
analysis (see Figure~\ref{whel_fit}). 

 \begin{center}
 \begin{tabular}{|l|c|c|c|c|}
\hline
 Sample  & SVX tags   & SLT tags   & No-tag & Dilepton \\ \hline
  Events &  34        &   14       &  46    & 7        \\
 Background & 9.2$\pm$1.2& 6.0$\pm$1.2 & 25.9$\pm$6.5& 0.76 $\pm$0.21 \\ \hline
\end{tabular}
\end{center}
Each lepton in a dilepton event is used, thus the total sample has 
108 leptons. 
The background in each momentum bin has different composition: 
b-mistag (non-top QCD), non-W
background, W$\bbbar$, etc. Background rates and background shapes 
are obtained from data and Monte Carlo. 

A maximum likelihood method is used to estimate F$_0$. The likelihood uses
the response functions for the different W helicity states and 
the background, and
constrains the background rate to the expected value. The input curves, 
the estimated background plus signal and the data are shown in 
Figure~\ref{whel_fit}. The result of the likelihood gives: 
\begin{eqnarray}
  F_0  = 0.91 \pm 0.37(stat) \pm 0.12(syst)  
\end{eqnarray}
where F$_0$ = 0.70 is expected.
The systematic error includes uncertainties from: M$_{top}$, acceptance
at low P$_T$, non-W background shape, gluon radiation in Monte Carlo and other
small contributions. The measurement is clearly dominated by the statistical
uncertainty.

 To evaluate a possible F$_+$ contribution, the F$_0$ value is fixed in
the likelihood to the SM prediction and a new maximization
is performed. The result  is:
\begin{eqnarray}
   F_+  = 0.11 \pm 0.15   
\end{eqnarray}
or F$_+ <$ 0.28 at 95\% CL, assuming  F$_0$ = 0.70.
CDF concludes that  with the present statistics no deviation from the SM is
observed.

\subsection{ Studies of $\ttbar$ spin correlations (${\rm D\!\not\!\!O}$) }

The SM predicts a width of 1.6 GeV for the top quark, this means that
it decays before hadronization and depolarization. At the Tevatron 
$\ttbar$ pairs are produced via $\qqbar$ pairs 90\% of the times, i.e.
via a spin 1 gluon. Top quarks are not individually polarized but as a pair
their spins are correlated. In the top quark rest frame
the angular  distributions of the decay products with respect to the
top quark spin vector, are as follows:
\begin{eqnarray}
   {dN /d cos\theta \sim (1 + \alpha cos \theta)}  
\end{eqnarray}
where we expect: 
\begin{center}
\begin{tabular}{|c|c|}
\hline
particle         &~~ $\alpha$ \\ \hline
e$^+$ or d quark &~~  1     \\
$\nu$ or u quark &~~ -0.31  \\
W$^+$            &~~ 0.41   \\
 b               &~~ -0.41  \\ \hline
\end{tabular}
\end{center}
Charged leptons carry most of the information about the top spin.
For dilepton events, i.e. $ \ttbar \into  \ell^+ \ell^- + X $, the
 correlations between the two leptons can be written as:
\begin{eqnarray}
  {{{dN} \over {d(cos\theta_+)d(cos\theta_-)}} \sim 
(1 + \kappa cos\theta_+ cos\theta_-)}  
\end{eqnarray}
where  all the spin correlation information is now in $\kappa$.
Here $\theta$ is the angle 
with respect to the so called ``off diagonal axis'',  defined
in the off-diagonal basis introduced by
Mahon and Parke ~\cite{ma-park}. This choice enhances the correlation by 
a factor 2.    At the Tevatron the expectation is 
$\kappa \sim$ 0.84~~\cite{mahlon}.

\begin{figure}[htbp]
 \centerline{
 \epsfysize=3.0in
 \epsffile[5 5 525 525] {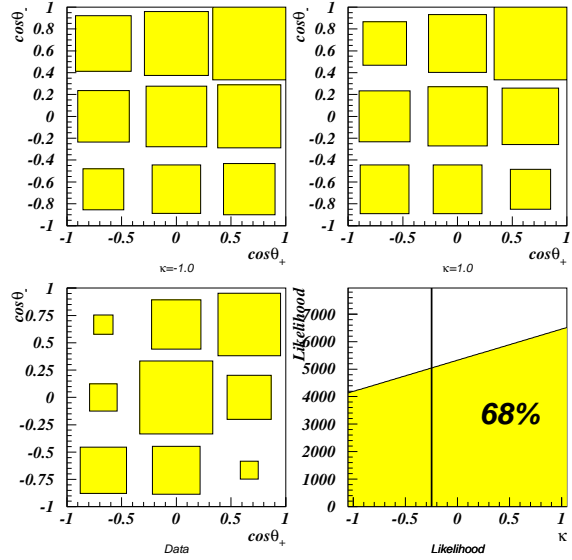  }}
 \caption{\label{d0-cor}${\rm D\!\not\!\!O}$ 
         spin correlation study. Top two plots are
  templates obtained from Monte Carlo for $\ttbar$ and background events. 
  The data is shown at bottom left, the results of the likelihood estimator
  are on the right. } 
\end{figure}

 ${\rm D\!\not\!\!O}$ uses the top dilepton sample of Run I, consisting of six events,
to check this correlation. 
Event reconstruction is needed to allow moving from one reference frame
to another. Because of missing constraints in the fitting
of dilepton events (two $\nu$ are missing) 
the sensitivity is somewhat diminished. Also combinatorics, due to 
multiple $\nu$ and $b$ assignments, and backgrounds
will dilute the expected effect.
A binned 2-dimensional likelihood is used to estimate
the value of $\kappa$. Templates, obtained from Monte Carlo for $\ttbar$
and for the background, are used for
values of $\kappa$ = -1.0 and 1.0. A total of 9 bins are 
used in Figure~\ref{d0-cor}. 

 ${\rm D\!\not\!\!O}$'s preliminary result is:
\begin{eqnarray}
  \kappa > -0.2~ \rm{~at ~the~} 68\% ~C.L.  
\end{eqnarray}     
Run II data will allow a good measurement of this correlation. 

%

\section{  Beyond the Standard Model }
 CDF has performed searches for rare top decays and a search for a state 
X $\into \ttbar$ that could indicate new strong dynamics. We summarize these
studies here.

\subsection{Search for rare Top Decays (CDF) } 
The large mass of the top quark suggests a strong connection with
the electroweak symmetry breaking sector. Any indication of unusual
top quark decays could shed some light on the mechanism for EWSB.
CDF has investigated two FCNC (Flavor Changing Neutral Currents) top 
decays: $t \into Z+q$ 
and $t \into \gamma +q $~~\cite{t_rare}. Both are expected 
by the SM to have branching ratios in the 10$^{-10}$ range.

  \underline{Search for $t \rightarrow Z q$ } 
Assuming FCNC the SM predicts
    BR($t \rightarrow Z q$) $\leq$ 0.001~~\cite{zq_ref}.
 CDF has searched in the channels ($\ell$ = $e$ or $\mu$)
\begin{eqnarray}
 \ppbar \into \ttbar \into Zq (W/Z)b \into \ell^+ \ell^- + 4~jets \\
\end{eqnarray}

  with jet E$_T >$ 20 GeV.
They find one Z $\rightarrow \mu \mu$ event.
 The background (from WZ, ZZ, ZZ pair production) is expected to be 
       B = 0.6 $\pm$ 0.2 events.
 Assuming that the observed event is signal and including all 
 systematic uncertainties, CDF sets a limit:
\begin{eqnarray}
  BR(t \rightarrow Z q) < 0.33  ~~~ (95\% ~CL)
\end{eqnarray}

\underline{Search for $t \rightarrow \gamma q$ }
 CDF search for $t \rightarrow \gamma q (q=u or c)$
 uses the production channel:
 \begin{eqnarray}
  q \bar{q} \rightarrow \ttbar \rightarrow W b \gamma q
 \end{eqnarray}
 The two relevant topologies are:
  \begin{enumerate}
   \item for W $\rightarrow \ell \nu$ they search for
      $ \gamma + \ell + \met + \ge$ 2 jets 
   \item for W $\rightarrow \qqbar \prime $ they search for
    $\gamma +\ge$4~jets and a $b$-tag 
  \end{enumerate}
  One event of type 1.) is observed. The expected background consists of 
  0.5 events from  W $\into \ell\nu$ and of 0.5 events from W $\into q q'$.  
 After inclusion of systematic uncertainties CDF sets a limit:
   \begin{eqnarray}
   BR(t \rightarrow \gamma q) < 0.032 ~~~~~(95\% ~CL) 
   \end{eqnarray}

\subsection{  Study of the $\ttbar$ Mass (CDF)   }

 New strong dynamics can be at the origin of a $\ttbar$ state.
 A number of models can be found in the literature on this subject.
In some models the $\ttbar$ state provides an explanation for
      the large value of M$_{top}$, in others top itself helps explain the 
      origin of mass.

As mentioned in
 Section~\ref{top_pt} the Topcolor model of Hill and Parke~\cite{top_color}
 or Topcolor assisted Technicolor model of Hill~\cite{topc-tech} ,  predict
 a process like:
\begin{eqnarray}
      \qqbar \into ~V~ \into \ttbar 
\end{eqnarray}
    where $V$ are color-octet (and/or singlet) Vector bosons.

 The multiscale Technicolor model of Eichten and Lane~\cite{techni} 
 predicts the process:
\begin{eqnarray}
     g~g \into ~\eta_T~ \into \ttbar 
\end{eqnarray}
   where the techni-eta is a scalar.
 In some part of the parameter space these states are narrow and can
 be detected. 

%
\begin{figure}[htbp]
 \centerline{
 \epsfysize=2.8in
 \epsffile[15 15 290 220] {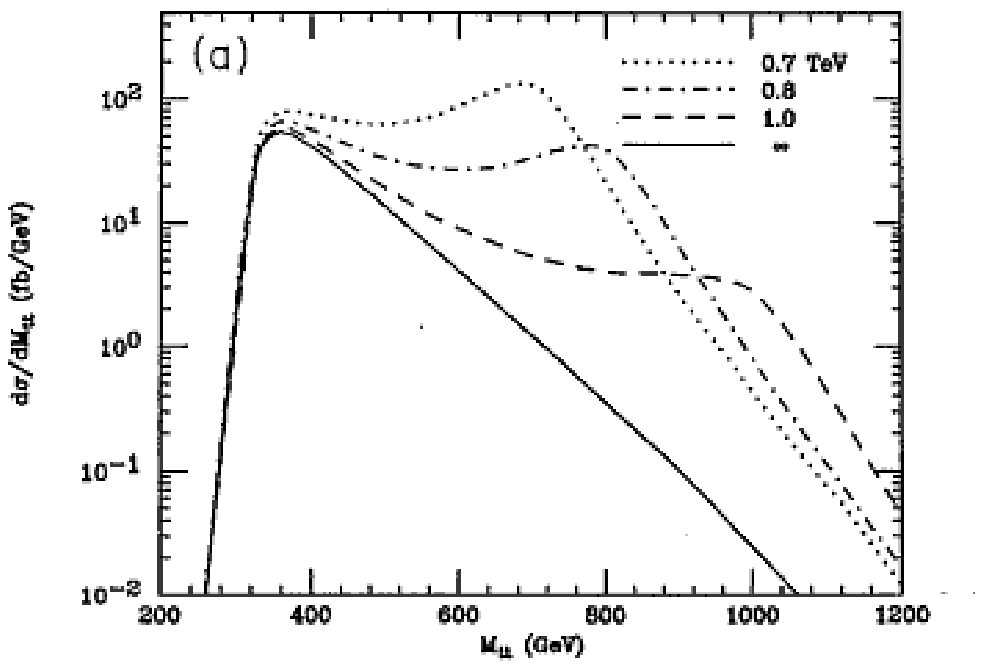 }
 \epsfysize=2.8in
 \epsffile[50 149 540 675] {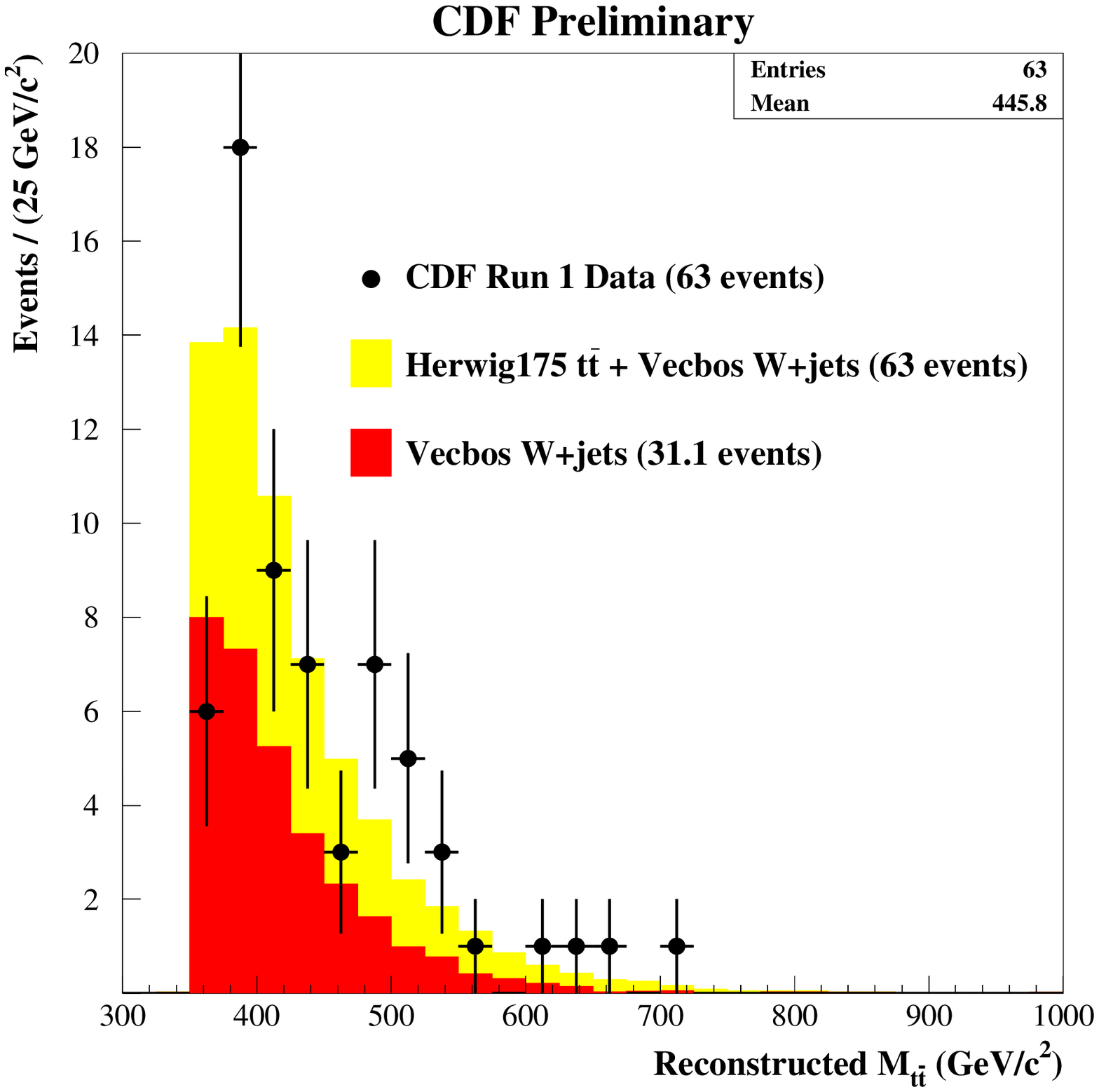 }}
 \caption{\label{mttbar} Left: M$_{\ttbar}$ distribution expectation from 
  a Topcolor model[12] with a narrow color singlet Z' $\into \ttbar$. 
  Right: distribution observed by CDF. }
\end{figure}

 CDF is looking at Run I data for possible hints of a $\ttbar$
 state and to establish cross section limits for these models.
 Event selection is similar to that used  for the top mass analysis: 
\begin{itemize}
 \item E$_T$(lepton) $>$  20 GeV; ~~~~~ $\met > $ 20 GeV 
 \item Raw E$_T$(jet 1-3) $>$ 15 GeV    
 \item Jet4: either tagged as $b$ or E$_T >$ 15 GeV 
\end{itemize}
The strategy used follows the mass analysis, i.e., the mass fitting 
kinematic constraint for event reconstruction is used with M$_{top}$ fixed at 
      175 GeV/c$^2$ (3C fit).  The goodness of fit, $\chi^2$ cut, is loosened
 to 50 (usually $\chi^2 <$ 10 is required for a 2C fit). 
To reduce effects due to wrong jet combinations they require 
    that the 3-body 
   masses in each event (for the same combination that gave the best fit)
   be in the interval 150-200 GeV/c$^2$ when computed using the measured
 quantities (rather than the result of the fit).

A total of 63 events remain with a background of 31 events. 
The expected M$_{\ttbar}$ distributions for background and background +
signal are shown in Figure~\ref{mttbar} and compared with the data. 
No excess over the expectation is observed.
Using the Topcolor assisted Technicolor  model of Hill~\cite{topc-tech}, and 
including systematic uncertainties on the measurement, CDF evaluates 
limits on cross sections for
      production of a colored vector boson, a Z', as a function of its mass.
The mass limit for a Topcolor Z' with $\Gamma(Z')$ = 0.012
is found to be M(Z') $>$ 630 GeV/c$^2$.

This is physics to be pursued in Run II.

\section{Summary and Conclusions}

A number of studies have been done at the Tevatron by CDF and ${\rm D\!\not\!\!O}$
to verify that top quark production is in agreement with  expectations
of the Standard model. No deviations from the SM have been found in studies
of top P$_T$ in $\ttbar$ production or in studies of top decay products and
spin correlations. In preparation for the next data taking period, Run II,
studies of single top production, search for rare top decays and for
a state X $\into \ttbar$ are being done. No new phenomena have been
observed.


\end{document}